\documentclass{iaus}
\usepackage{graphicx}

\title[The Galaxy and its stellar halo - a hybrid cosmological approach] {The
  Galaxy and its stellar halo - insights from a hybrid cosmological approach}

\author[De Lucia \& Helmi]   
{Gabriella De Lucia$^1$
 \and Amina Helmi$^2$}

\affiliation{$^1$Max--Planck--Institut f\"ur Astrophysik, \\
  Karl--Schwarzschild--Str. 1, D-85748 Garching, Germany \\
  email: {\tt gdelucia@mpa-garching.mpg.de} \\[\affilskip]
  $^2$Kapteyn Astronomical Institute, University of Groningen, 
  \\ P.O. Box 800, 9700 AV Groningen, Netherlands 
  \\email: {\tt ahelmi@astro.rug.nl}}

\pubyear{2008}
\volume{254}  
\pagerange{1--6}
\jname{The Galaxy Disk in Cosmological Context}
\editors{J. Andersen, J. Bland-Hawthorn \& B. Nordstr\"om, eds.}
\begin{document}

\maketitle

\begin{abstract}
We use a series of high-resolution N-body simulations of a `Milky-Way' halo,
coupled to semi-analytic techniques, to study the formation of our own Galaxy
and of its stellar halo. Our model Milky Way galaxy is a relatively young
system whose physical properties are in quite good agreement with observational
determinations. In our model, the stellar halo is mainly formed from a few
massive satellites accreted early on during the galaxy's lifetime. The stars in
the halo do not exhibit any metallicity gradient, but higher metallicity stars
are more centrally concentrated than stars with lower abundances.  This is due
to the fact that the most massive satellites contributing to the stellar halo
are also more metal rich, and dynamical friction drags them closer to the inner
regions of the host halo. 
\end{abstract}

\firstsection 
\section{Introduction}

Our own galaxy - the Milky Way - is a fairly large spiral galaxy consisting of
four main stellar components: (1) the thin disk, that contains most of the
stars with a wide range of ages and on high angular momentum orbits; (2) the
thick disk, that contains about 10-20 per cent of the mass in the thin disk
and whose stars are on average older and have lower metallicity than those in
the thin disk; (3) the bulge, which contains old and metal rich stars on low
angular momentum orbits; and (4) the stellar halo which contains only a few per
cent of the total stellar mass and whose stars are old and metal poor and
reside on low angular momentum orbits.

While the Milky Way is only one Galaxy, it is the one that we can study in
unique detail. Over the past years, accurate measurements of ages,
metallicities and kinematics have been collected for a large number of
individual stars, and much larger datasets will become available in the next
future thanks to a number of ongoing and planned astrometric, photometric and
spectroscopic surveys. This wealth of detailed and high-quality observational
data provides an important benchmark for current theories of galaxy formation
and evolution.

In the following, we outline the main results of a recent study of the
formation of the Milky Way and of its stellar halo in the context of a hybrid
cosmological approach which combines high-resolution simulations of a `Milky
Way' halo with semi-analytic methods. We refer to \cite[De Lucia \& Helmi
(2008)]{DeLucia_Helmi_2008} for a more detailed description of our method and
of our results.

\section{The simulations and the galaxy formation model}

We use the re-simulations of a `Milky-Way' halo (the GA series) described in
\cite[Stoehr et al. (2002)]{Stoehr_etal_2002} and \cite[Stoehr et
al. (2003)]{Stoehr_etal_2003}, with an underlying flat $\Lambda$-dominated CDM
cosmological model. The candidate halo for re-simulations was selected from an
intermediate-resolution simulation (particle mass $\sim 10^8\,{\rm M}_{\odot}$)
as a relatively isolated halo which suffered its last major merger at
$z>2$. The same halo was then re-simulated at four progressively higher
resolution simulations with particle mass $\sim 1.7\times10^8$ (GA0), $\sim
1.8\times 10^7$ (GA1), $\sim 1.9\times 10^6$ (GA2), and $\sim 2.1\times
10^5\,{\rm M}_{\odot}$ (GA3). Simulation data were stored in 108 outputs from
$z=37.6$ to $z=0$, and for each simulation output we constructed group
catalogues (using a standard friends-of-friends algorithm) and substructure
catalogues (using the {\small SUBFIND} algorithm developed by \cite[Springel et
al. 2001]{Springel_etal_2001}). Substructure catalogues were then used to
construct merger history trees for all self-bound haloes as described in
\cite[Springel et al. (2005)]{Springel_etal_2005} and \cite[De Lucia \& Blaizot
(2007)]{DeLucia_Blaizot_2007}. Finally, these merger trees were used as input
for our semi-analytic model of galaxy formation. 

\section{Physical properties and metallicity distributions}

\begin{figure}[b]
  \begin{center}
    \includegraphics[width=4.8in]{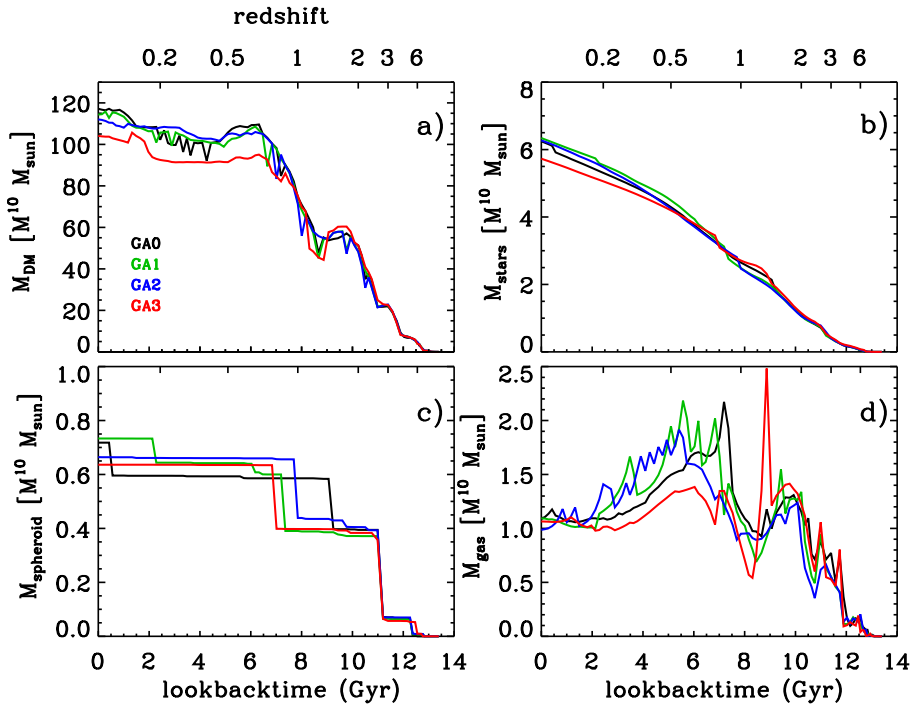} 
    \caption{Evolution of the dark matter mass (panel a), total stellar mass
      (panel b), spheroid mass (panel c), and cold gas mass (panel d) for the
      model Milky Way in the four simulations used in our study (different
      colours).}
    \label{fig1}
  \end{center}
\end{figure}

Fig.\,\ref{fig1} shows the evolution of different mass components for the model
Milky Way galaxies in the four simulations used in our study (lines of
different colours). The histories shown in the different panels have been
obtained by linking the galaxy at each time-step to the progenitor with the
largest stellar mass. Fig.\,\ref{fig1} shows that approximately half of the
final mass in the dark matter halo is already in place (in the main progenitor)
at $z\sim 1.2$ (panel a) while about half of the final total stellar mass is
only in place at $z\sim 0.8$ (panel b). About 20 per cent of this stellar mass
is already in a spheroidal component (panel c). The mass of the spheroidal
component grows in discrete steps as a consequence of our assumption that it
grows during mergers and disk instability episodes, and approximately half of
its final mass is already in a spheroidal component at $z\sim 2.5$. In
contrast, the cold gas mass varies much more gradually. 

Interestingly, the model produces consistent evolutions for all four
simulations used in our study, despite the large increase in numerical
resolution. Some panels (e.g. panel b) do not show perfect convergence, due to
the lack of complete convergence in the N-body simulations (see panel
a). Fig.\,\ref{fig1} also shows that the total stellar mass of our model Galaxy
($6\times 10^{10}\,{\rm M}_{\odot}$) is is very good agreement with the
estimated value $\sim 5-8\times 10^{10}\,{\rm M}_{\odot}$. The mass of the
spheroidal component is instead slightly lower than the observed value (assumed
to be about 25 per cent of the disk stellar mass), while our fiducial model
gives a gas mass which is about twice the estimated value.

\begin{figure}[h]
  \begin{center}
    \includegraphics[width=5.2in]{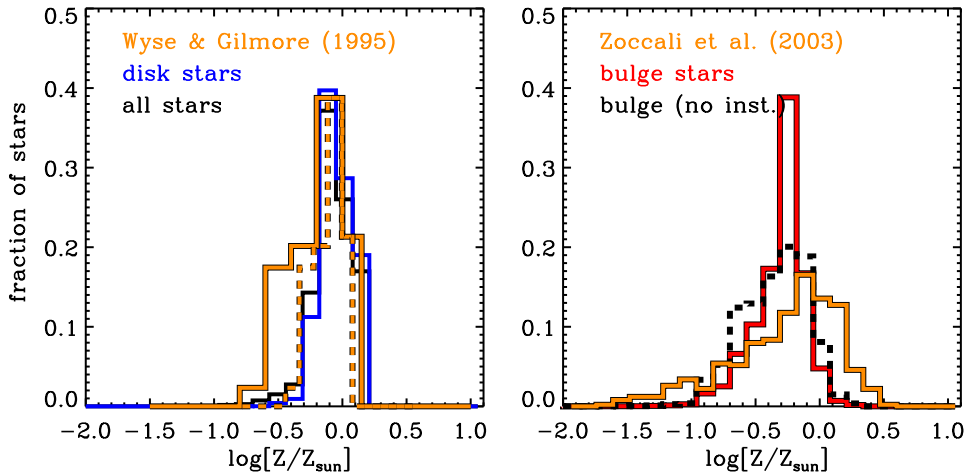} 
    \caption{Metallicity distribution for stars in the disk (blue histogram in
    the left panel) and spheroid (red histogram in the right panel) of the
    model Milky Way from the highest resolution simulation in our study. The
    solid black histogram in the left panel shows the metallicity distribution
    for all stars in the model galaxy, while the dashed black histogram in the
    right panel shows the metallicity distribution of stars in the spheroidal
    component for our fiducial model if spheroid growth through disk
    instability is suppressed. The solid orange histograms show observational
    measurements by Wyse \& Gilmore (1995, left panel) and Zoccali et
    al. (2003, right panel). The dashed orange histogram in the left panel has
    been obtained converting the [Fe/H] scale of the original distribution by
    Wyse \& Gilmore into an [O/H] scale by using the observed [O/H]-[Fe/H]
    relation for thin disk stars by Bensby, Feltzing \& Lundstr\"om (2004).}
    \label{fig2}
  \end{center}
\end{figure}

Fig.\,\ref{fig2} shows the metallicity distributions of the stars in the disk
and spheroid of our model Milky Way from the highest resolution simulation used
in our study. The left panel shows the metallicity distribution of all stars
(black) and of the stars in the disk (blue) compared to the observational
measurements by \cite[Wyse \& Gilmore (1995)]{Wyse_Gilmore_1995}. The right
panel shows the metallicity distribution of the spheroid stars in our fiducial
model (red) and in a model where the disk instability channel is switched off
(dashed black). Model results are compared to observational measurements by
\cite[Zoccali et al. (2003)]{Zoccali_etal_2003}. The metallicity distribution
of disk stars in our model peaks at approximately the same value as observed,
but it exhibits a deficiency of low metallicity stars. When comparing model
results and observational measurements, however, two factors should be
considered: (1) the observational measurements have some uncertainties ($\sim
0.2$~dex) which tend to broaden the true underlying distributions; (2) the
observational measurements provide {\it iron} distributions, and iron is not
well described by our model that adopts an instantaneous recycling
approximation. In order to show the importance of this second caveat we have
converted the measured [Fe/H] into [O/H] using a linear relation, obtained by
fitting data for thin disk stars from \cite[Bensby, Feltzing \& Lundstr\"om
(2004)]{Bensby_etal_2004}. The result of this conversion is shown by the dashed
orange histogram in the left panel of Fig.~\ref{fig2}. The observed [O/H]
metallicity distribution is now much closer to the modelled log[Z/Z$_\odot$]
distribution. The same caveats applies to the comparison shown in the right
panel, which indicates that our model spheroid is significantly less metal rich
than the observed Galactic bulge.

\section{The stellar halo}

In order to study the structure and metallicity distribution of the stellar
halo, we assume that it builds up from the cores of the satellite galaxies that
merged with the Milky Way over its lifetime. The stars that end up in the
stellar halo are identified by tracing back all galaxies that merged with the
Milky Way progenitor, until they are central galaxies of their own halo. We
select then a fixed fraction (10 per cent for the results shown in the
following) of the most bound particles of their parent haloes, and tag them
with the mean metallicity of the central galaxies (for details, see De Lucia \&
Helmi 2008). 

\begin{figure}[h]
  \begin{center}
    \includegraphics[width=5.2in]{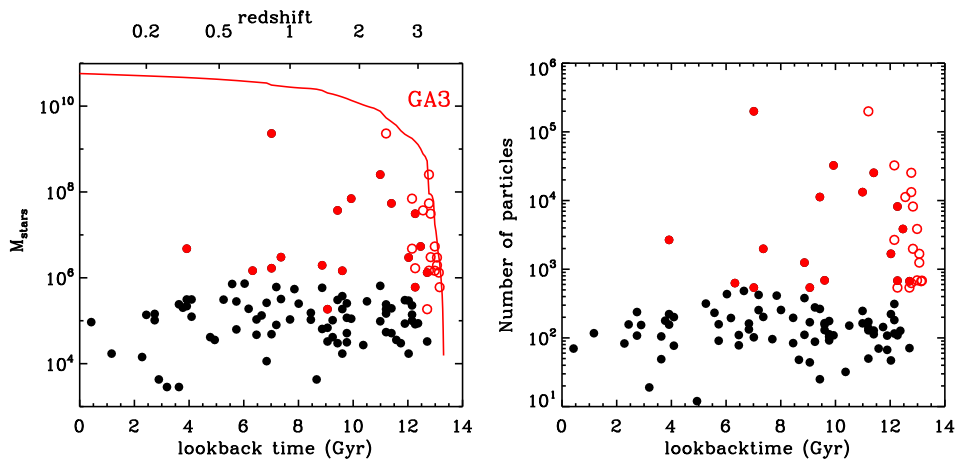}
    \caption{Stellar mass (left panel) of the galaxies contributing to the
      stellar halo, as a function of the lookback time of galaxy's merger. The
      right panel shows the number of particles associated to the dark matter
      haloes at the time of accretion. Red symbols correspond to objects with
      more than 500 particles. Open symbols correspond to red filled circles
      but are plotted as a function of the time of accretion.}
    \label{fig3}
  \end{center}
\end{figure}

\begin{figure}[t]
  \begin{center}
    \includegraphics[width=2.6in]{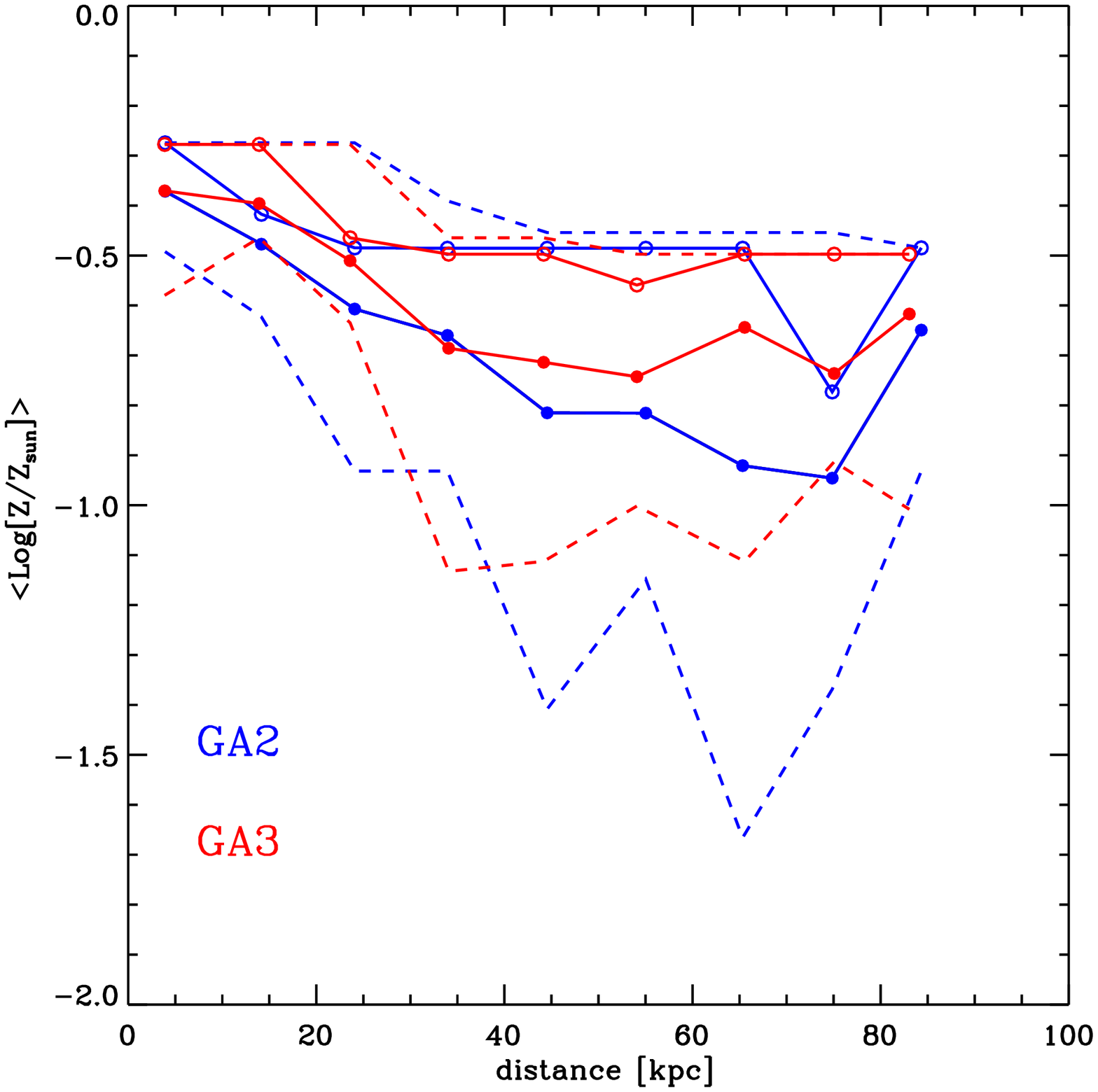} 
    \includegraphics[width=2.6in]{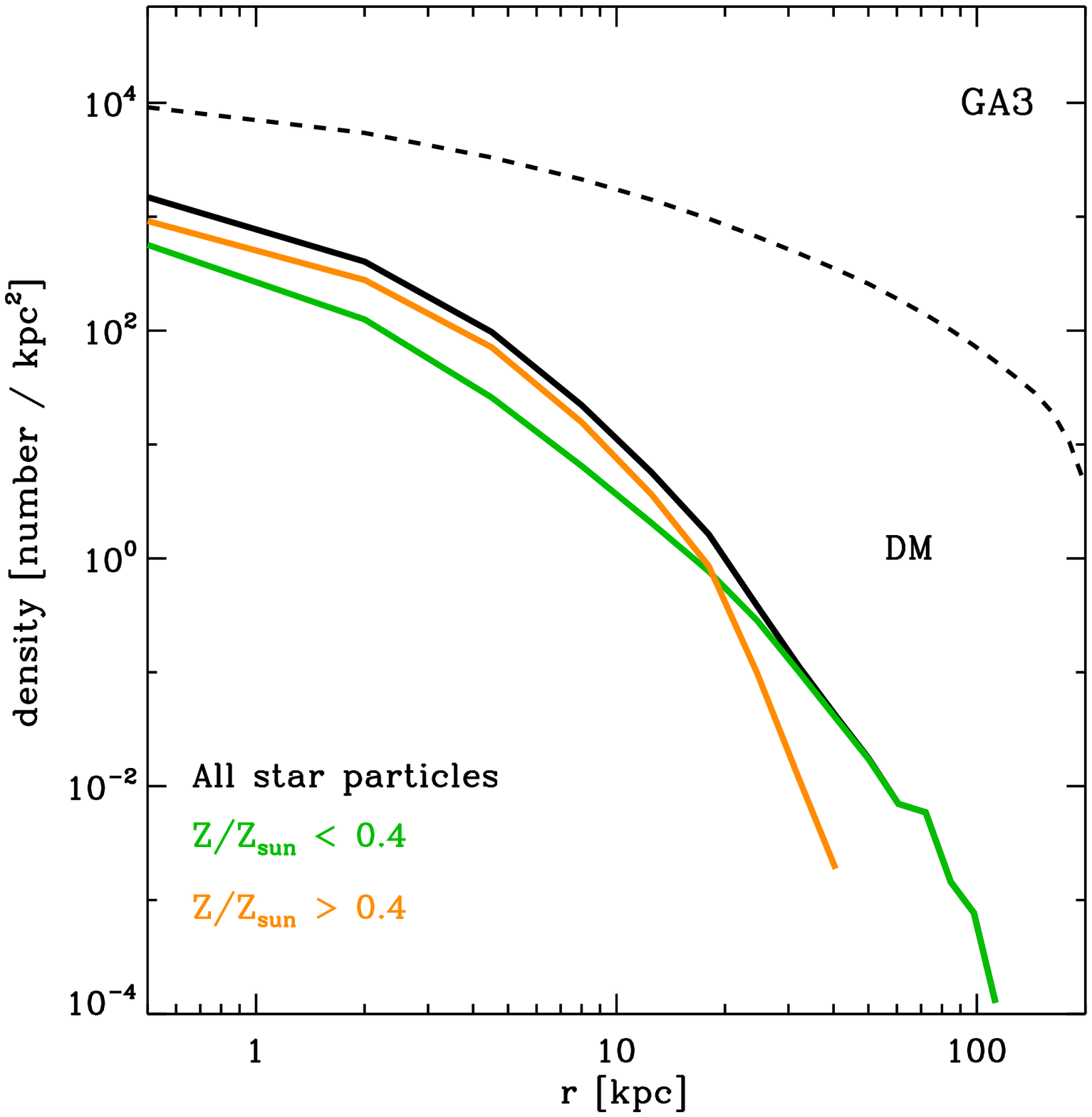} 
    \caption{Left panel: Mean (filled circles) and median (empty circles)
      metallicity of star particles as a function of the distance from the most
      bound particle in the Milky Way halo, for the simulations GA2 (blue) and
      GA3 (red). Dashed lines correspond to the 15th and 85th percentiles of
      the distribution. Right panel: Projected density profile of the stellar
      halo (solid black line) and of the dark matter halo (dashed black line)
      for the simulation GA3. The solid green and orange lines show the
      projected density profiles for star particles with metallicity smaller
      and larger than $0.4\,Z_{\odot}$ respectively.}
    \label{fig4}
  \end{center}
\end{figure}

Fig.\,\ref{fig4} shows the stellar mass of the galaxies contributing to the
stellar halo as a function of the lookback time of the galaxy's merger (left
panel), and the number of particles associated to the dark matter haloes at the
time of accretion (right panel). Most of the accreted galaxies lie in quite
small haloes and only a handful of them are attached to relatively more massive
systems, which are accreted early on during the galaxy's lifetime. These are
the galaxies that contribute most to the build-up of the stellar halo. The
results illustrated in Fig.\,\ref{fig4} are in good agreement with those by
\cite[Font et al. (2006)]{Font_etal_2006} who combined mass accretion histories
of galaxy-size haloes with a chemical evolution model for each accreted
satellites to study the formation of the stellar halo.

Fig.\,\ref{fig4} shows the metallicity of star particles as a function of
the distance from the most bound particle in the Milky Way halo for the
simulations GA2 (blue) and GA3 (red). For both simulations, the mean
metallicity decreases from ${\rm Log[Z/Z}_{\odot}{\rm]}\sim -0.4$ at the centre
to $\sim -0.8$ at a distance of $\sim 40\,{\rm kpc}$. The median and upper 85th
percentile of both distributions are approximately flat around $\sim -0.5$.
Note that the metallicity of our stellar halo is higher than what is known for
the Galactic halo near the Sun. We note also that both distributions are
dominated in number by star particles associated to one or a few accreted
galaxies with relatively high metallicity (hence the flat behaviour of the
median and upper percentile of the distribution). The lower percentile declines
with increasing distance from the centre, suggesting that the inner region is
largely dominated by high-metallicity stars while the contribution from lower
metallicity stars becomes more important moving to the outer regions. This is
shown more explicitely in the right panel of Fig.\,\ref{fig4} which shows the
projected density profile of the stellar halo (black) for the simulation
GA3. The solid orange and green lines in this panel show the projected profiles
of star particles with metallicity larger and smaller than $0.4\,Z_{\odot}$
respectively. High metallicity stars are more centrally concentrated than stars
with lower abundances, suggesting that the probability of observing
low-metallicity stars increases at larger distances from the Galactic centre
($\gtrsim 10-20$~kpc), where the contribution from the inner more metal-rich
stars is less dominant. Interestingly, this result appears to be in qualitative
agreement with recent measurements by \cite[Carollo et
al. (2007)]{Carollo_etal_2007}. 

In our model, the `dual' nature of the stellar halo originates from a
correlation between the stellar metallicity and the stellar mass of accreted
galaxies. Since the most massive galaxies decay through dynamical friction to
the inner regions of the halo, this is where higher metallicity stars will be
found preferentially.

\section{Conclusions}

We have combined high-resolution resimulations of a `Milky Way' halo with
semi-analytic techniques to study the formation of our own Galaxy and of its
stellar halo. The galaxy formation model used in our study has been used in a
number of previous studies and has been shown to provide a reasonable agreement
with a large number of observational data both in the local Universe and at
higher redshifts (De Lucia \& Helmi 2008 and references therein). Our study
demonstrates that the same model is able to reproduce quite well the observed
physical properties of our own Galaxy. The agreement is not perfect: our model
Galaxy contains about twice the gas observed in the Milky Way, and the model
bulge is slightly less massive and substantially less metal rich than the
Galactic bulge. A detailed comparison between model results and observational
measurements of metallicity distributions is complicated by the use of an
instantaneous recycling approximation which is not appropriate for the
iron-peak elements, mainly produced by supernovae Type Ia. Relaxing of this
approximation in future work, will allow us to carry out a more detailed
comparison with observed chemical compositions, and to establish similarities
and differences between present-day satellites and the building blocks of the
stellar halo.

Our model stellar halo is made up of very old stars (older than $\sim 11$~Gyr)
with low metallicity, although higher than what is known for the stellar halo
of our Galaxy. Most of the stars in the halo are contributed by a few
relatively massive satellites accreted early on during the galaxy's lifetime.
The building blocks of the stellar halo lie on a well defined mass-metallicity
relation. Since the most massive galaxies are dragged closer to the inner
regions of the halo by dynamical friction, this produces a stronger
concentration of more metal rich stars, in qualitative agreement with recent
observational measurements. The numerical resolution of our simulations is too
low for the study of spatially and kinematically coherent structures in model
stellar halo. Higher resolution simulations are needed for this kind of study.


\begin{thebibliography}{}

\bibitem[Bensby, Feltzing \& Lundstr\"om (2004)]{Bensby_etal_2004}
{Bensby, T., Feltzing, S. \& Lundstr\"om, I.} 2004, 
\textit{A\&A}, 415, 155 

\bibitem[Carollo \etal\ (2007)]{Carollo_etal_2007}
{Carollo, D., Beers, T. C., Lee, Y. S., Chiba, M., Norris, J. E., Wilhelm, R.,
  Silvarani, T., Marsteller, B., Munn, J. A., Bailer-Jones, C. A. L.,
  Fiorentin, P. R. \& York, D. G.} 2007,  
\textit{Nature}, 450, 1020 

\bibitem[De Lucia \& Blaizot (2007)]{DeLucia_Blaizot_2007}
{De Lucia, G. \& Blaizot, J.} 2007, 
\textit{MNRAS}, 375, 2 

\bibitem[De Lucia \& Helmi (2008)]{DeLucia_Helmi_2008}
{De Lucia, G. \& Helmi, A.} 2008, 
\textit{MNRAS} submitted, arXiv:0804.2465 

\bibitem[Font \etal\ (2006)]{Font_etal_2006}
{Font, A. S., Johnston K. V., Bullock J. S. \& Robertson, B. E.} 2006, 
\textit{ApJ}, 638, 585

\bibitem[Springel \etal\ (2001)]{Springel_etal_2001}
{Springel, V., White, S. D. M., Tormen, G., \& Kauffmann, G.} 2001, 
\textit{MNRAS}, 328, 726 

\bibitem[Springel \etal\ (2005)]{Springel_etal_2005}
{Springel, V., White, S. D. M., Jenkins, A., Frenk, C. S., Yoshida, N., Gao,
  L., Navarro, J., Thacker, R., Croton, D., Helly, J., Peacock, J. A., Cole,
  S., Thomas, P., Couchman, H., Evrard, A., Corlberg, J. \& Pearce, F.} 2005, 
\textit{Nature}, 435, 629 

\bibitem[Stoehr \etal\ (2002)]{Stoehr_etal_2002}
{Stoehr, F., White, S. D. M., Tormen, G., \& Springel, V.} 2002, 
\textit{MNRAS}, 335, L84 

\bibitem[Stoehr \etal\ (2003)]{Stoehr_etal_2003}
{Stoehr, F., White, S. D. M., Springel, V., Tormen, G. \& Yoshida, N.} 2003, 
\textit{MNRAS}, 345, 1313 

\bibitem[Wyse \& Gilmore (1995)]{Wyse_Gilmore_1995}
{Wyse, R. F. G. \& Gilmore, G.} 1995, 
\textit{AJ}, 110, 2771 

\bibitem[Zoccali \etal\ (2003)]{Zoccali_etal_2003}
{Zoccali, M., Renzini, A., Ortolani, S., Greggio, L., Saviane, I., Cassisi, S.,
  Rejkuba, M., Barbuy, B., Rich, R. M. \& Bica, E.} 2003, 
\textit{A\&A}, 399, 931

\end{thebibliography}
\end{document}